\documentclass[prx,aps,twocolumn,amsmath,amssymb]{revtex4}

\pdfoutput=1
\usepackage{hyperref}
\usepackage{graphicx}
\usepackage[usenames]{color}
\usepackage{bm}

\definecolor{blue}{rgb}{0,0,1}

\usepackage[usenames]{color}
\newcommand{\new}[1]{\textcolor{black}{#1}}

\def\eqq#1{Eq.~(\ref{#1})}
\def\eq#1{(\ref{#1})}

\def\f#1{Fig.~\ref{#1}}

\def\c#1{~\cite{#1}}
\def\cc#1{Ref.\c{#1}}

\def\s#1{Section~\ref{#1}}
\def\kt{k_{\rm B}T}
\def\e{{\rm e}}

\def\beq{\begin{equation}}
\def\eeq{\end{equation}}
\def\bea{\begin{eqnarray}}
\def\eea{\end{eqnarray}}

\begin{document}

\title{Varied phenomenology of models displaying dynamical large-deviation singularities}
\author{Stephen Whitelam$^1$}
\email{swhitelam@lbl.gov}
\author{Daniel Jacobson$^2$}
\email{daniel.jacobson@caltech.edu}

\affiliation{$^1$Molecular Foundry, Lawrence Berkeley National Laboratory, 1 Cyclotron Road, Berkeley, CA 94720, USA\\
$^2$Division of Chemistry and Chemical Engineering, California Institute of Technology, Pasadena, California 91125, USA}

\begin{abstract}
  Singularities of dynamical large-deviation functions are often interpreted as the signal of a dynamical phase transition and the coexistence of distinct dynamical phases, by analogy with the correspondence between singularities of free energies and equilibrium phase behavior. Here we study models of driven random walkers on a lattice. These models display large-deviation singularities in the limit of large lattice size, but the extent to which each model's phenomenology resembles a phase transition depends on the details of the driving. We also compare the behavior of ergodic and non-ergodic models that present large-deviation singularities. We argue that dynamical large-deviation singularities indicate the divergence of a model timescale, but not necessarily one associated with cooperative behavior or the existence of distinct phases.
\end{abstract}

\maketitle

\section{Introduction}

Phase transitions are collective phenomena that occur in the limit of large system size and whose presence can be detected in finite systems\c{goldenfeld2018lectures,binney1992theory,chandler1987introduction}. Phase transitions cause singularities in thermodynamic potentials and dynamical large-deviation functions, which quantify the logarithmic probability of observing particular values of extensive order parameters\c{den2008large,touchette2009large,gallavotti1995dynamical,lebowitz1999gallavotti}. \new{An important example of this singularity-phase coexistence correspondence in equilibrium} is the 2D Ising model below its critical temperature\c{chandler1987introduction,binney1992theory,goldenfeld2018lectures,onsager1944crystal}. In dynamical models, singularities (kinks) of large-deviation functions develop in certain limits, and can signal the emergence of a dynamical phase transition and the coexistence of distinct dynamical phases\c{corberi2019probability,garrahan2009first,vaikuntanathan2014dynamic,pietzonka2016extreme,nemoto2017finite,nyawo2017minimal, klymko2017similarity,klymko2017rare,horowitz2017stochastic,nyawo2018dynamical,nemoto2019optimizing,mallmin2019comparison,jack2020ergodicity,casert2020dynamical}. 

However, large-deviation singularities do not necessarily indicate the existence of cooperative phenomena or distinct phases. For instance, singular features are seen in the large-deviation functions of finite systems in the reducible limit, when the connections between microstates are severed\c{dinwoodie1992large,dinwoodie1993identifying,coghi2019large,garrahan2014comment,whitelam2018large}. We show here that singularities can also appear in the limit of large system size of dynamical models, if the model's basic timescale (mixing time) diverges with system size. Such singularities appear whether or not this divergence results from cooperative behavior or is accompanied by evidence of distinct phases.

We study models of driven random walkers on a lattice, which display dynamical large-deviation singularities in the limit of large system size. If the walker is driven in one direction then we see the emergence of dynamical intermittency within trajectories conditioned to produce particular values of a dynamical order parameter. The switching time of this intermittency grows with system size. If the walker is undriven, the singularity results instead from a divergence of the diffusive mixing time of the model, with no intermittency present in conditioned trajectories (both behaviors have a thermodynamic realization in terms of a lattice polymer). We present an argument to rationalize when to expect random-walk models to exhibit intermittency of their conditioned trajectory ensembles, and show that this argument correctly predicts the mixed intermittent/non-intermittent character of the conditioned trajectory ensemble of a random walker whose driving varies with position. We also comment on the relationship between ergodic and non-ergodic dynamical systems that exhibit large-deviation singularities.

In \s{section_driven} and \s{section_undriven} we consider two random-walk models that display large-deviation singularities, but whose conditioned trajectory ensembles are of different character. In \s{sec_discussion} we present a simple argument to rationalize when such models display intermittency of their conditioned trajectories. In \s{grow} we compare the behavior of ergodic and non-ergodic dynamical models that present large-deviation singularities. We conclude in \s{sec_conc}, arguing that dynamical large-deviation singularities indicate the divergence of a model timescale, but not necessarily one associated with cooperative behavior or the existence of distinct phases.

\section{Driven random walker}\label{section_driven}

\begin{figure*}
   \centering
   \includegraphics[width=\linewidth]{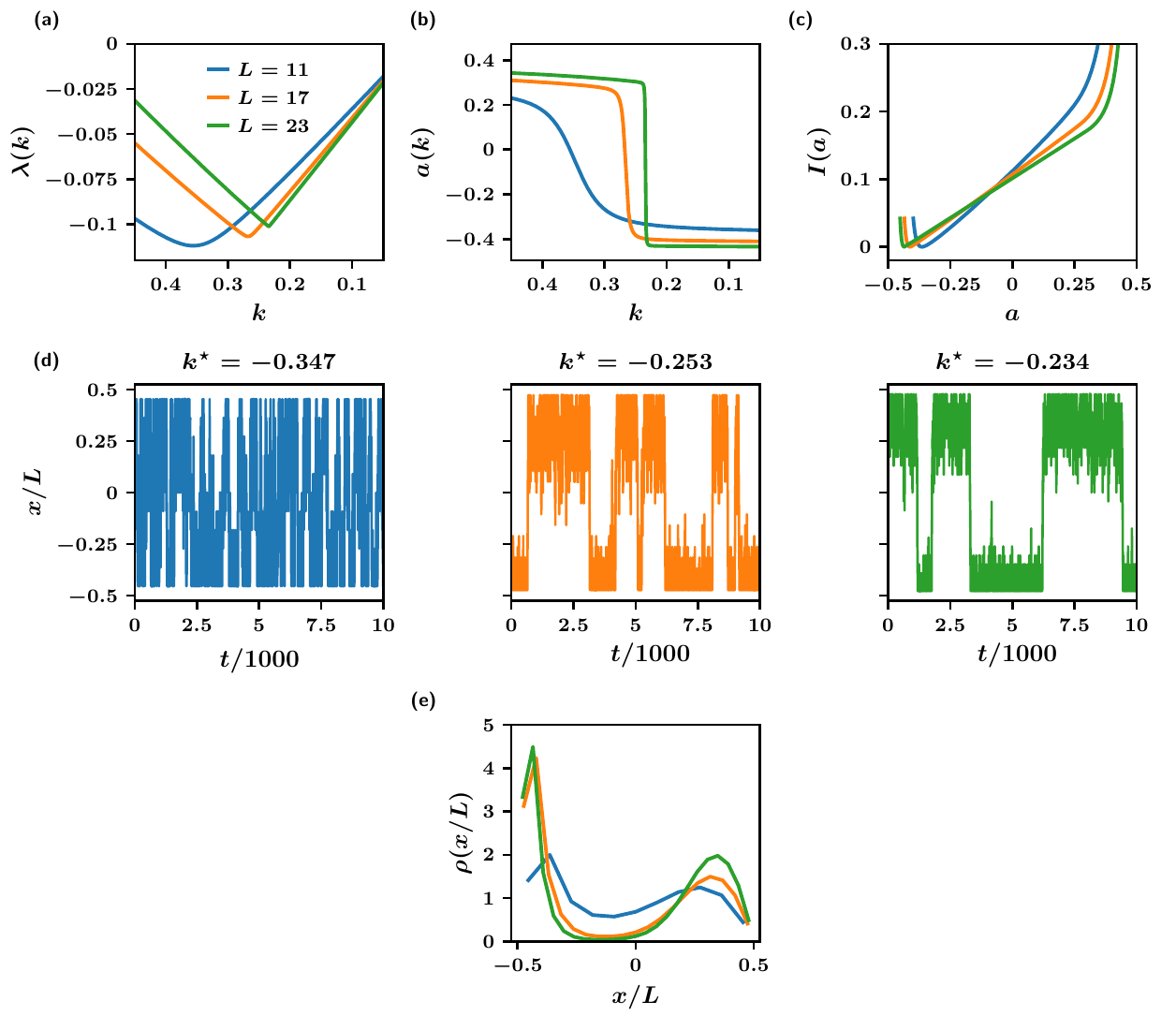} 
   \caption{(a--c) Large-deviation functions for the time-averaged position $a$ of a driven discrete random walker on a closed lattice of $L$ sites. (d) Walker trajectories showing the instantaneous position $x/L$ under the biased dynamics corresponding to the point $k=k^\star$ of greatest curvature of $\lambda(k)$. These trajectories show intermittency, with the walker switching between two locations on either side of the lattice. The timescale for residence in the distinct lattice locations increases with increasing $L$. (e) Histograms of the instantaneous position $x/L$ for the trajectories in (d).}
\label{fig_driven}
\end{figure*}

We start with a model similar to one studied in~\cc{nemoto2017finite}, a driven random walker on a closed \new{(non-periodic)} lattice of $L$ sites. We choose $L$ to be odd, and work in discrete time~\footnote{We have carried out analogous calculations in continuous time, and draw the same conclusions.}. Let the instantaneous position of the walker be $x \in \{-(L-1)/2, \dots, (L - 1)/2\}$. At each time $t$ the walker moves right $(x \to x+1)$ with probability $p(x)$, or left with probability $1-p(x)$. In this section we set $p(x) =1/4$, and so the walker's typical location is near the left-hand side of the lattice, $x=-(L-1)/2$. If the walker sits at either edge of the lattice then it moves away from the edge with probability $1$ (so $p(-(L-1)/2) =1$ and $p((L-1) / 2) = 0$), \new{analogous to reflecting boundaries in the continuum limit.}

The master equation associated with this dynamics is
\begin{equation}
  P_x(t+1) = \sum_{x'} W_{x'x} P_{x'}(t),
\end{equation}
where $P_x(t)$ is the probability that the walker resides at lattice site $x$ at time $t$, and the generator $W_{x'x}=p(x') \delta_{x,x'+1} + (1-p(x')) \delta_{x,x'-1}$ is the probability of the transition $x' \to x$.
\begin{figure*}
   \centering
  \includegraphics[width=\linewidth]{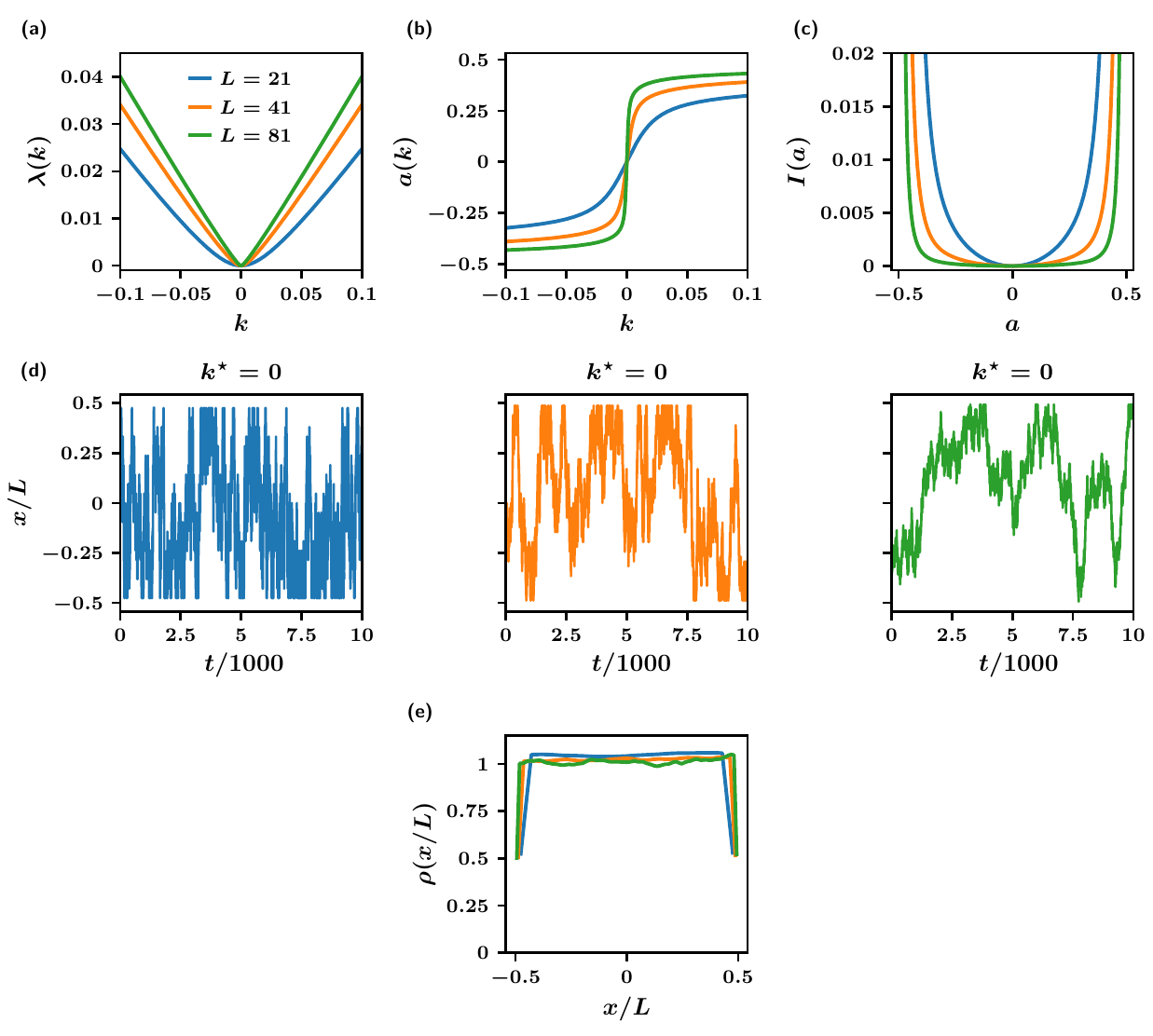} 
   \caption{Analog of \f{fig_driven}, now for an undriven walker. (a--c) As for the driven walker, large-deviation functions show increasingly sharp behavior as $L$ increases. (d) Trajectories showing the instantaneous position $x/L$ of the walker at the point of greatest curvature of $\lambda(k)$, $k^\star = 0$. These trajectories do not exhibit intermittency. (e) As a result, histograms of the instantaneous position $x/L$ for the trajectories in (d) are unimodal.}
\label{fig_unbiased}
\end{figure*}
\begin{figure*}
   \centering
   \includegraphics{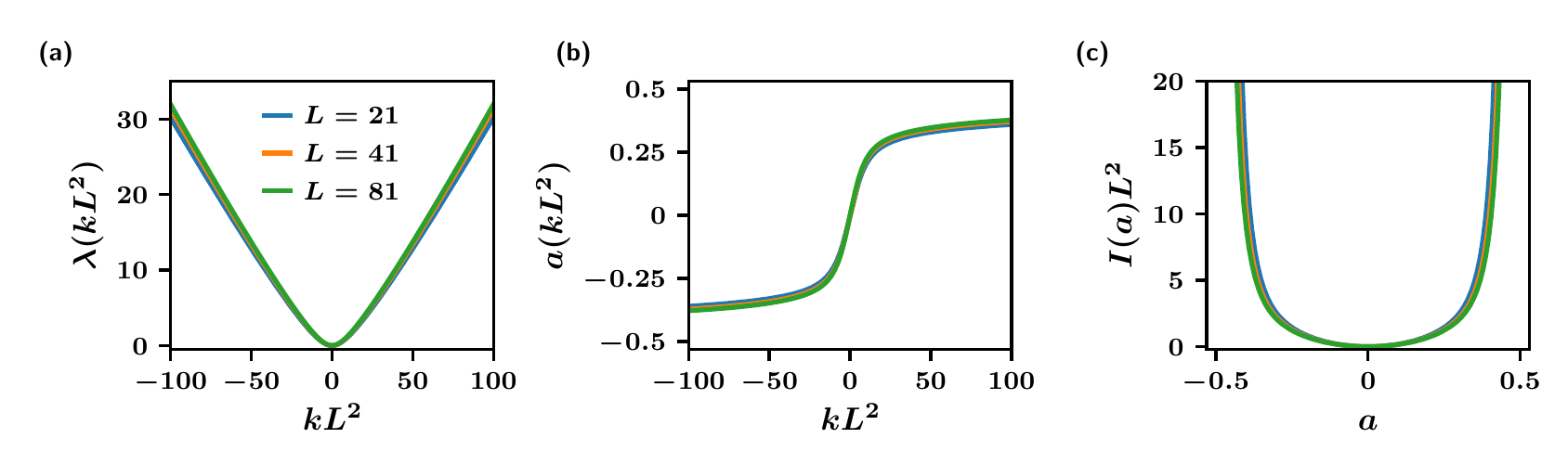} 
   \caption{(a--c) The large-deviation functions of \f{fig_unbiased} (a--c), rescaled by $L^2$ to account for the timescale associated with diffusion. The resulting collapse indicates that these systems behave similarly when viewed on the natural timescale $T/L^2$. The large-deviation singularity in this case results from divergence of the diffusive timescale.}
   \label{fig_unbiased_rescaled}
\end{figure*}

We take the time-averaged position $a$ of the walker as our dynamical observable. \new{This quantity is}
\begin{equation}\label{one}
	a(\omega)=(TL)^{-1} \sum_{t=1}^T x^\omega_t,
\end{equation}
\new{where $x^\omega_t$ is the position of the walker at time $t = 1,\dots, T$ within a trajectory $\omega$.} We have normalized $a(\omega)$ by the size of the lattice, $L$. \new{The typical value of $a$, which we call $a_0$, corresponds to the value of \eq{one} in the limit of large $T$. Because the walker prefers to sit near the left-hand side of the lattice, $a_0 \approx -1/2$.}

To calculate the probability distribution $\rho_T(A=aT)$ of the walker's time-averaged position, we appeal to the tools of large-deviation theory. The probability distribution adopts in the long-time limit the large-deviation form
\begin{equation}\label{dynamical_LDP}
\rho_T(A) \approx e^{-T I(a)},
\end{equation}
where $I(a)$ is the rate function (on speed $T$)\c{den2008large,touchette2009large}. $I(a)$ quantifies the probability with which the walker achieves a specific, and potentially rare, time-averaged position. When $I(a)$ is convex, as it is for ergodic Markov chains, it can be recovered from its Legendre transform, the scaled cumulant-generating function (SCGF)\c{touchette2009large},
\begin{equation}\label{scgf}
  \lambda(k) = a(k) k - I(a(k)).
\end{equation}
Here $k$ is a conjugate field, and $a(k)=\lambda'(k)$ is the value of $a$ associated with a particular value of $k$. If the lattice is not too large then the SCGF can be calculated by finding directly the largest eigenvalue of the tilted generator, $W^k_{x'x} =\e^{k x} W_{x'x}$. The rate function can then be obtained by inverting \eq{scgf}. We use this standard method to calculate $\lambda(k)$, $a(k)$, and $I(a)$.

In \f{fig_driven} we show the large-deviation functions for the time-averaged position $a$ of the driven walker. As the lattice size $L$ increases, the SCGF and $a(k)$ bend increasingly sharply, and portions of the rate function become increasingly linear.

In \f{fig_driven}(d) we show biased dynamical trajectories of the walker, generated at the points $k = k^{\star}$ at which the SCGF bends most sharply. We generated these trajectories using the exact eigenvectors of the tilted generator\c{chetrite2015variational,ray2018exact}. Because the SCGF is convex, biased trajectories generated using field $k$ correspond to trajectories that produce a value $a(k)=\lambda'(k)$ of the time-integrated observable $a$\c{chetrite2015variational}, \new{and are the ``least unlikely of all the unlikely ways''\c{den2008large} of achieving the specified time average. For this model these trajectories} are intermittent, with the walker switching abruptly from one location on the lattice to another. As a result, histograms of the instantaneous position of the walker are bimodal [panel (e)]. As the lattice size increases, the residence time at each location increases.

The intermittent behavior has a simple physical origin. The probability per unit time for the walker to sit at (fluctuate about) its preferred location is greater than that to sit at any site in the lattice interior, but the latter probability is essentially independent of position (see \s{sec_discussion}). If conditioned to achieve a time-averaged position $a$ at (say) the center of the lattice, it could sit for all time at the corresponding lattice location. But it could also spend half its time at its preferred location, and half its time near the far end of the lattice. Given that sitting near the far end of the lattice is not more costly than sitting in the middle, the intermittent strategy is more probable than the homogeneous one. This argument holds for time $T$ much longer than $\tau(L)$, the emergent mixing time governing intermittency. The probability of crossing the lattice in the difficult direction is $\sim p^L$, and so the timescale for doing so increases exponentially with $L$. 

\section{Undriven random walker}
\label{section_undriven}

We now consider an undriven walker whose probability of moving right at any site away from the edges is $p(x)=1/2$. Again we choose the time-averaged position of the walker as our dynamical observable. As shown in \f{fig_unbiased}, the large-deviation functions $\lambda(k)$ and $a(k)$ again show the emergence of sharp features as $L$ grows, and $I(a)$ flattens, reminiscent of the free energy for the Ising model below its critical temperature\c{binney1992theory}. These sharp features become singular in the limit $L \to \infty$, with the kink occurring at $k^\star=0$ (the untilted generator for a random walker has a spectral gap that vanishes as $L^{-2}$).

However, the implication of the emergence of distinct ``phases'' or dynamic intermittency is at odds with the physics of the system. Because of the model's symmetry, the point $k=k^\star$ at which the SCGF shows greatest curvature is $k^\star=0$, corresponding to the unbiased trajectory ensemble. Such trajectories do not display switching behavior, as shown in panels (d) and (e). We also verified that switching behavior occurs at no other values of $k$: histograms of $x/L$ for biased or conditioned trajectory ensembles are always peaked about a single value. Why then the emergent singularity?

To answer this question we note that the rate function for dynamics \new{controls the rate} at which atypical fluctuations decay, and so measures both the probability and basic timescale of those fluctuations. Thus $I(a)$ can be small if the fluctuation $a$ is almost typical, or if the basic timescale governing the establishment and decay of a fluctuation $a$ is large. For diffusive systems such as the walker, the latter factor is important. The natural way to compare systems of different $L$ is at fixed scaled observation time $T_L \equiv T/L^2$, which in large-deviation terms is equivalent to adopting $T_L$ as the new large-deviation speed, such that $\rho_T(A) \approx \e^{-T I(a)} = \e^{-T_L I_L(a)}$.
The object $I_L(a) \equiv I(a) L^2$ is the rate function on this new speed. 

In \f{fig_unbiased_rescaled} we show the large-deviation functions for the walker in this new frame of reference. Each panel is a rescaling of the panels shown in \f{fig_unbiased}(a--c). These rescaled functions show no sharpening of their features as $L$ increases, and the collapse of the functions confirms that the long timescale associated with the walker is the diffusive one. The natural scales for comparison of these systems is $T/L^2$ and $k L^2$, not $T$ and $k$.

Therefore in this case it is a divergence of the diffusive mixing time $L^2$ that causes the singularity, not the emergence of intermittent behavior. 
One additional issue resolved by the rescaling is the apparent vanishing of the rate function in the limit $L\to \infty$. If a large-deviation principle applies then the rate function $I(a)$ has a unique zero at the point $a_0$ at which the system displays its typical behavior\c{touchette2009large}. Given the symmetry of the system, the walker's typical location in the long-time limit is $a_0 = 0$. It is clear that \f{fig_unbiased_rescaled}(c) is consistent with this idea, and the notion that time is ``long''.

In Appendix~\ref{thermo} we point out that both walker models have a thermodynamic interpretation as lattice polymers, confirming that the existence of a first-order singularity in a thermodynamic system does not automatically imply phase coexistence.
  \begin{figure}
   \centering
  \includegraphics[width=\linewidth]{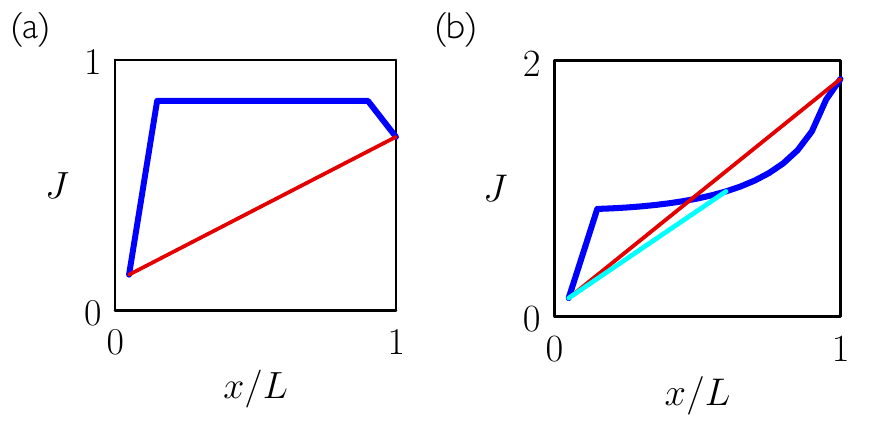} 
   \caption{Negative log-probability per unit time for the driven walker to achieve a time-averaged position $a=x/L$ in a homogeneous way (blue lines) or a two-state intermittent way (cyan and red lines). In panel (a) the walker's driving is constant, $p(x)=1/4$, while in panel (b) the tendency of the walker to move left increases with rightward distance, $p(x) = [1-(x/L)^2]/4$. The lattice size $L=20$. When the straight lines lie below the blue lines it is more likely for the system to achieve the time-averaged position $a$ corresponding to $x/L$ in an intermittent way. This construction is consistent with conditioned trajectories of these models (\f{fig_vard_walkers}), as long as the trajectory time comfortably exceeds the time on which switching occurs.}
   \label{fig_potential}
\end{figure}

\section{When should we expect intermittency?}
\label{sec_discussion}

The previous sections show that intermittent conditioned trajectories can accompany dynamical large-deviation singularities, but that singularities can result from the emergence of a large timescale {\em absent} intermittency. We show in this section that driven walkers on a lattice can display both intermittent and non-intermittent conditioned dynamics, depending upon the the details of the walker rules and the timescale of observation. \new{The argument we use is analogous to the classic equilibrium procedure of comparing the free energies of homogenous and coexisting phases\c{goldenfeld2018lectures,binney1992theory}.}

Consider again a lattice of $L$ sites, and let the probability that a walker steps right from lattice site $x\in\{1,\dots,L\}$ be $p(x)$ (we have shifted the origin of the lattice relative to the previous sections). Define $q(x)\equiv 1-p(x)$. Let the lattice be closed, so that $p(1) = 1$ and $p(L)=0$. The time-integrated position of the walker is $a = (LT)^{-1} \sum_{t=1}^T x_t$, where $x_t$ is the walker's position at discrete time $t$. 

\begin{figure}
   \centering
  \includegraphics[width=0.7\linewidth]{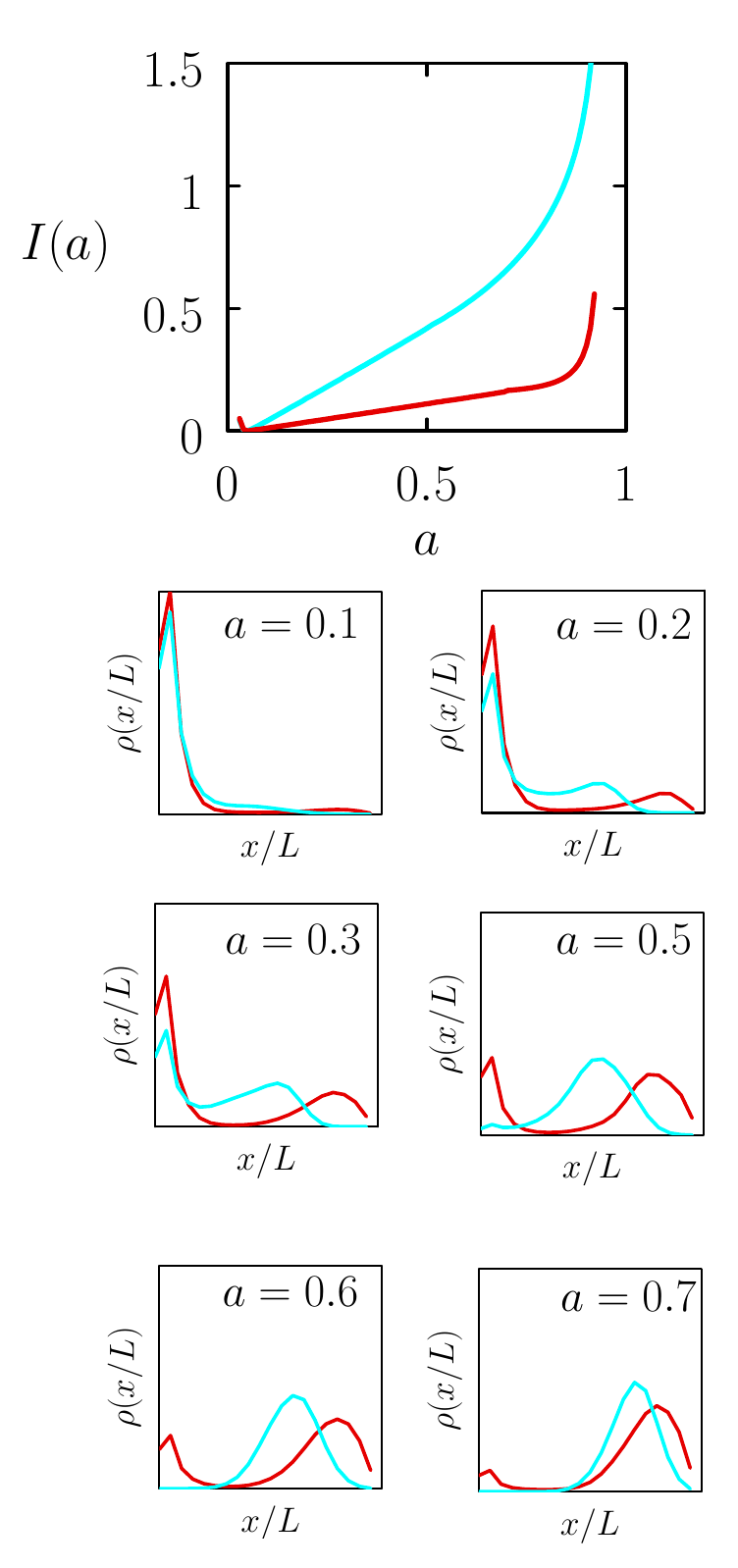} 
   \caption{Large-deviation rate function $I(a)$ for the time-averaged position of the driven lattice random walker from \s{section_driven} (red), together with that for a super-driven walker (cyan) for which the probability to move left increases with distance from the left-hand edge. Shown below are histograms of the instantaneous walker position from trajectories of the two models conditioned to produce various atypical values of $a$. Consistent with the simple arguments developed in this section, the driven walker displays intermittency, for most values of $a$, involving sites near the edges of the lattice. By contrast, the super-driven walker shows intermittency for $a$ not too far from the typical value $a_0$, involving sites near the middle of the lattice. For $a$ far from $a_0$, its conditioned trajectories are homogeneous. For both models, $L=20$.}
   \label{fig_vard_walkers}
\end{figure}

\new{The probability for a walker at an interior site $x$ to fluctuate about that site, i.e. to step right, left, left and then right again is $p(x) q(x+1) q(x) p(x-1)$. Thus the negative logarithmic probability per unit time for the walker to remain localized near interior site $x$ is the negative logarithm of this quantity divided by 4. Accounting for the different rates at the edges of the lattice, the negative logarithmic probability per unit time for the walker to remain localized near site $x$ is}
\beq
U(x) = 
\begin{cases}
-\frac{1}{2} \ln [p(x)q(x+1)] & x=1\\
-\frac{1}{4} \ln [p(x)q(x+1)q(x)p(x-1)]& 1<x<L\\
-\frac{1}{2} \ln [q(x)p(x-1)] & x=L.
\end{cases}
\eeq
 The negative logarithmic probability per unit time for a homogeneous trajectory, one localized at $x=aL$ for all time, is 
\beq
\label{homog}
J_{\rm homog}= U(x).
\eeq 
(The quantity $J$ is not the rate function $I(a)$, which relates to the log-probability by which the system achieves the value $a$ by any means.) 

By contrast, the negative logarithmic probability per unit time for an intermittent trajectory built from sections of trajectory localized at $x'=a'L$ for time $\phi T$ and at $x''=a''L$ for time $(1-\phi)T$ is 
\beq
\label{int}
J_{\rm int} = \phi U(x')+(1-\phi) U(x''),
\eeq
ignoring switches back and forth between $x'$ and $x''$. For the intermittent trajectory to achieve the same time average as the homogeneous one requires 
\beq
\label{balance}
\phi x' + (1-\phi) x''=x.
\eeq
If, for a given value of $x=aL$, \eq{int} is smaller than \eq{homog}, then intermittent trajectories are more probable than homogeneous trajectories. There may be other types of trajectory that are more probable still, but this simple and approximate argument, which essentially reduces to an assessment of where $U(x)$ is concave, provides a starting point for understanding when intermittency will appear in the conditioned dynamics of the walker.

In \f{fig_potential}(a) we show the quantities \eq{homog} and \eq{int} for the driven walker of \s{section_driven}, for which $p(x) = 1/4$. The negative log-probability per unit time for homogeneous trajectories, \eqq{homog}, is shown in blue. The most probable intermittent trajectory for any $x$ is the one built from the lattice edges, shown in red in the figure. This line lies below the homogeneous result for all $x$ away from the edges, showing that intermittency is the more probable way to achieve a time average in the interior of the lattice.

In \f{fig_potential}(b) we consider a ``super-driven'' walker whose probability of moving in one direction increases with distance in the opposite direction, $p(x) = [1-(x/L)^2]/4$. For intermittent trajectories to be more likely than homogeneous ones we need to be able to draw a straight line between two points on the function $U(x)$ and have the line lie below $U(x)$. We see that this is possible for some points but not others, and that the lower-lying line (the more probable intermittent strategy) connects the typical point $x \approx 1$ with a point that is near the middle of the lattice, not the edge. Based on this picture we expect the conditioned trajectory ensemble of the super-driven walker to be intermittent for values of $a$ near the typical value $a_0 \approx 1/L$, but not far away, and for the intermittent trajectories to involve locations on the lattice different to the edges occupied by the driven walker.

In \f{fig_vard_walkers} we show that these expectations are borne out. In the main panel we show the rate functions $I(a)$ for the time-averaged position of the driven walker (red) and super-driven walker (cyan). Shown below are position histograms for trajectories conditioned to produce various atypical values of $a$. Consistent with the simple arguments developed in this section, the super-driven walker shows intermittency near $a=a_0$, involving sites near the center of the lattice. For $a$ far from $a_0$ its conditioned trajectories are homogeneous. The driven walker shows intermittency for a wider range of values of $a$, and its intermittency always involves sites near the edges of the lattice.

To produce \f{fig_vard_walkers} we used the VARD method\c{jacobson2019direct} to calculate the conditioned dynamics of the walker. The unconditioned model has probability $p(x)$ of moving right from lattice site $x$. We introduce a reference random walker whose probability of moving right, $\tilde{p}(x)$, is an arbitrary function of $x$, which we chose to express as a radial basis function neural network. The network has one input node, which takes the value $x/L$, a single hidden layer of 1000 neurons, each with Gaussian activations, and one output node, $\tilde{p}(x)$. For sufficiently long trajectories the optimal dynamics is Markovian, and can be represented exactly by this ansatz if suitably optimized.

Following~\cc{whitelam2020evolutionary} we used neuroevolution of the parameters of the network, equivalent to gradient descent in the presence of Gaussian white noise\c{whitelam2020correspondence}, to extremize the sum of values of $T^{-1} \ln [\tilde{p}(x)/p(x)]$ over a trajectory of $T$ steps of the reference random walker's dynamics, subject to its achieving a specified value of the time-averaged position $a$. For $T$ long enough, i.e. longer than any emergent mixing time of the reference model, these calculations are equivalent to the eigenvalue calculations of \s{section_driven} and \s{section_undriven}\c{chetrite2013nonequilibrium}, and we recover the rate function of the model and its conditioned dynamics at each point on the rate function.

For $T/\tau(L)$ not large, conditioned trajectories do not resemble their long-time counterparts. For example, for $T=L$ the most probable trajectory whose time-averaged location is the middle of the box, given free choice of initial conditions, is the one that starts at the right-hand wall and crosses the box in $L$ steps. The histogram $\rho(x/L)$ associated with that trajectory is flat. For $T \ll L$ the only viable trajectories respecting the conditioning are those localized near the appropriate value of $x/L$. 

\section{Singularities in non-ergodic models}
\label{grow}
\begin{figure}[b] 
   \centering
  \includegraphics[width=\linewidth]{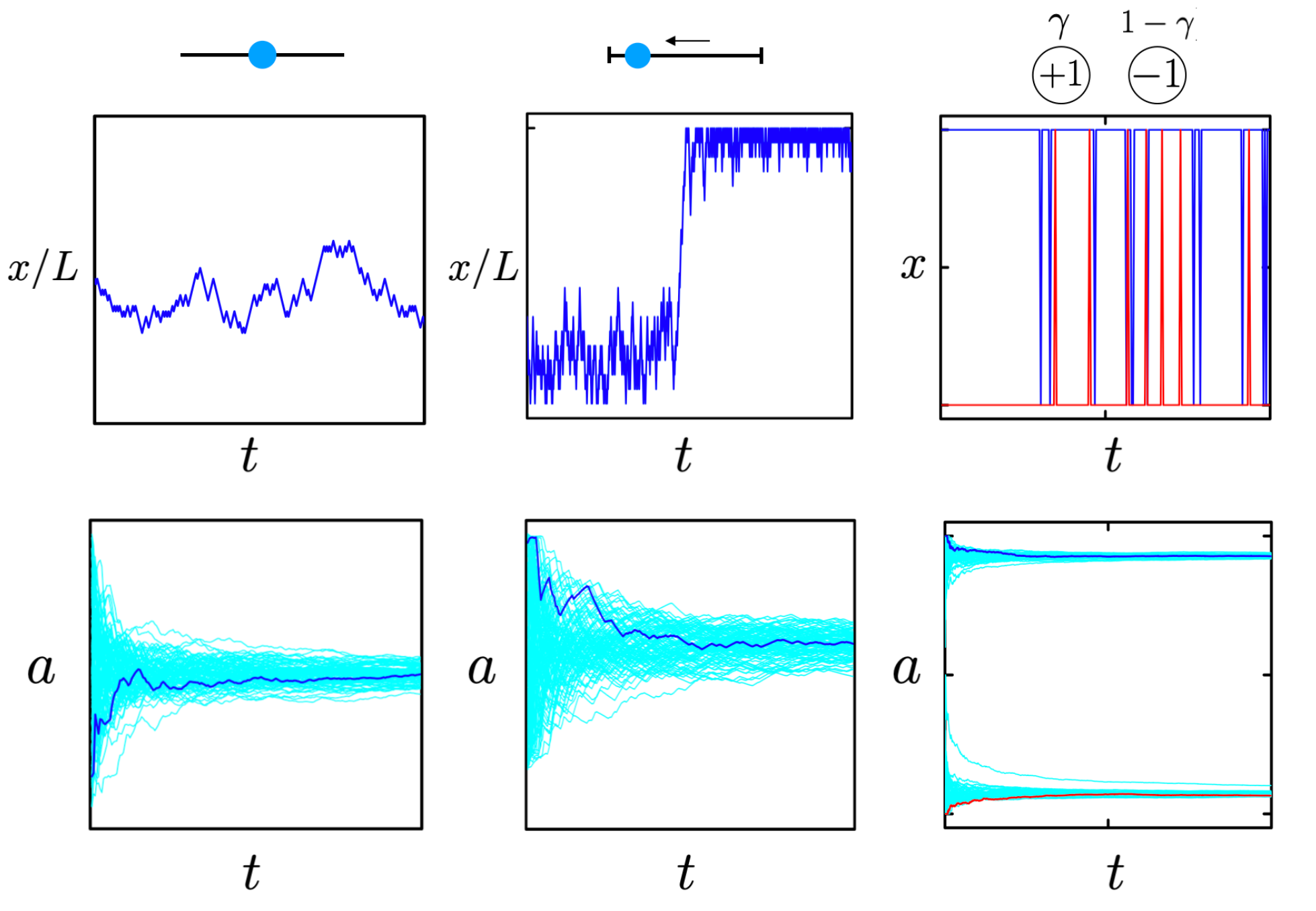} 
   \caption{Summary of the biased trajectory ensembles associated with three model systems, each generated at the point $k^\star$ at which the SCGF associated with the time-integrated observable has largest curvature. For the undriven walker (left) and the growth model (right) we have $k^\star=0$, while for the driven walker (center) we have $k^\star \neq 0$. The top panels show the instantaneous dynamical observable associated with a single trajectory (two trajectories in the case of the growth model). The lower panels show $a$, the time-integrated version of $x$, for an ensemble of 100 trajectories. The trajectories of the top panels are indicated in the bottom panels by the dark blue or red lines.}
\label{fig_comparison}
\end{figure}

We end by noting that dynamical large-deviation singularities also arise in non-ergodic models, such as the irreversible growth model of Refs.\c{klymko2017similarity,klymko2017rare}, and that these singularities are associated with phenomenology of a distinct type to that exhibited by the walker models.

Briefly, the growth model possesses two types of particle $x \in \{-1,1\}$, added to an urn at discrete timesteps with a relative probability that depends on the quantity $\gamma(a) = \e^{-Ja}/(2 \cosh J a)$, where $J$ is a parameter. Here $a$, the dynamical observable, is the sum of values of $x$ at each timestep, divided by total time $T$. This model can also be viewed as a two-state switch with a switching probability that depends on the history of switching\c{whitelam2018large}.

This model undergoes a phase transition, at a value of $J=1$, between a regime in which trajectories display one type of characteristic behavior and a regime in which trajectories display two types of characteristic behavior. At the critical point the trajectory ensemble displays anomalous fluctuations. Associated with this transition is a change of shape of the model's large-deviation rate function (see Appendix~\ref{app_grow}), and a dynamical large-deviation singularity. 

Long trajectories of the conditioned driven walker and the growth model show two-state switching behavior, but in the growth model's case the probability of switching depends on the history of switching. As a result, trajectories that adopt one type of behavior become more likely, as time advances, to remain committed to that behavior\c{morris2014growth}. The result is ergodicity breaking and an ensemble of trajectories that in the long-time limit spontaneously adopt one of two characteristic behaviors. 

The nature of the trajectory ensemble of the growth model and walker models is summarized in \f{fig_comparison}. Conditioned trajectories of the driven walker display intermittency and a bimodal distribution of the {\em instantaneous} coordinate $x/L$, but the distribution of the {\em time-integrated} quantity $a$ is unimodal. In the growth model, the distribution of the time-integrated quantity $a$ is bimodal.

\section{Discussion \& Conclusions}
\label{sec_conc}

Phase transitions are collective phenomena that occur in the limit of large system size, and which influence the behavior of finite systems\c{goldenfeld2018lectures,binney1992theory,chandler1987introduction}. Phase transitions induce singularities in thermodynamic potentials and large-deviation functions. However, similar-looking singularities can arise in the absence of collective phenomena. For instance, abrupt features are seen in the large-deviation functions of finite systems in the reducible limit, when the connections between microstates are severed\c{dinwoodie1992large,dinwoodie1993identifying,coghi2019large,garrahan2014comment,whitelam2018large}. Here we have shown that singularities can also emerge in the limit of large system size if a model becomes slow as it becomes large, whether or not it exhibits behavior reminiscent of a phase transition. 

The undriven walker of \s{section_undriven} has a rate function $I(a)$ [\f{fig_unbiased}(c)] that in the limit $L \to \infty$ looks similar to that of the 2D Ising model's magnetization rate function below $T_{\rm c}$\c{binney1992theory,touchette2009large}. Interpreting the walker in this context suggests that it can be switched between two behaviors (corresponding to walkers localized either size of the lattice) with an infinitesimal field $k$. Moreover, because the switching occurs about the value $k^\star=0$, a large unbiased version of the system appears poised on the brink of phase coexistence between these behaviors. However, the conditioned trajectory ensemble of the walker shows no evidence of distinct dynamical phases. In our view a more natural interpretation is that the walker's diffusive timescale diverges as $L^2$. When time $T$ and field $k$ are rescaled in order to view systems of different size at fixed $T/L^2$, is clear that the probability distribution of $a$ remains regular (\f{fig_unbiased_rescaled}). Departures from the typical behavior (a time average in the middle of the lattice) remain rare. The distinction between large-deviation singularities induced by cooperative behavior or by dynamics that is simply slow is the idea expressed in Fig. 1 of \cc{whitelam2018large}.

The driven walker of \s{section_driven} displays emergent intermittency when its trajectories are conditioned to produce particular time-averaged positions, provided that $T \gg \tau(L)$, the emergent switching time. In \s{sec_discussion} we show that the details of the walker's rules and the value of the imposed time averaged determine whether intermittency or homogeneity dominates. Other authors have noted the similarity of dynamical intermittency to magnetization stripes in the Ising strip crystal, if we associate time with the long Ising box direction and regard the two walker lattice positions as ``phases''\c{nemoto2017finite}. However, if we swap the long and short directions of the Ising box then the direction of the interface changes, but the identity of the phases does not. If we do similarly with the walker, and make $L \gg T$, then its conditioned trajectories can no longer be intermittent. This change switches the direction of the walker's space-time ``interface'', but also alters the identity and number of the ``phases'' seen.

The growth model (or two-state switch with memory) discussed in \s{grow} is unlike the other models discussed in that it possesses only one independent dimension, that of time $T$. It is clearly different in detail to models with spatial degrees of freedom, but its phenomenology is similar to that of the 2D Ising model in several important respects, with $T$ playing the role of system size. The growth model's large-deviation function changes from being regular to being singular upon changing a model parameter $J$. In the singular regime the steady-state rate function of the time-extensive quantity is concave (\s{app_grow}), reflecting two distinct dynamical behaviors and an associated ergodicity breaking. The origin of this behavior is cooperativity, the tendency of the model to favor one behavior the more it exhibits it.

\new{The scaled cumulant-generating functions (SCGFs) of these models, which display kinks in certain limits, do not specify which type of dynamics the model exhibits. It is worth noting that nor do the rate functions to which they are Legendre dual. For the walker models, the SCGF kinks are related to portions of the rate function that are linear with zero gradient (the undriven walker) and with nonzero gradient (the driven walker). The latter is suggestive of intermittency, although linear rate functions also arise in other types of process, such as relaxation to an absorbing state\c{oono1989large,touchette2009large}. Moreover, simple switching models, which are by design intermittent, display, as the switching time increases, rate functions that broaden and vanish, becoming linear with zero gradient\c{dinwoodie1992large,dinwoodie1993identifying,coghi2019large} (see e.g. Fig. 5 of~\cc{whitelam2018large}). Those rate functions therefore resemble the rate functions of the {\em undriven} walker, although the latter shows no intermittency. Explicit calculation of the dynamics that gives rise to the rate function is necessary to determine how the model realizes its rare behavior.}

All of the models discussed here display kinks of their dynamical large-deviation functions in certain parameter regimes, but show phase transition-like behavior to different extents. On this basis we suggest that phenomenology should guide the classification of singularity-bearing models discussed in the literature.

\section{Acknowledgments}
S.W. performed work at the Molecular Foundry, Lawrence Berkeley National Laboratory, supported by the Office of Science, Office of Basic Energy Sciences, of the U.S. Department of Energy under Contract No. DE-AC02--05CH11231.

\appendix
\begin{figure*}[t] 
   \centering
  \includegraphics[width=0.85\linewidth]{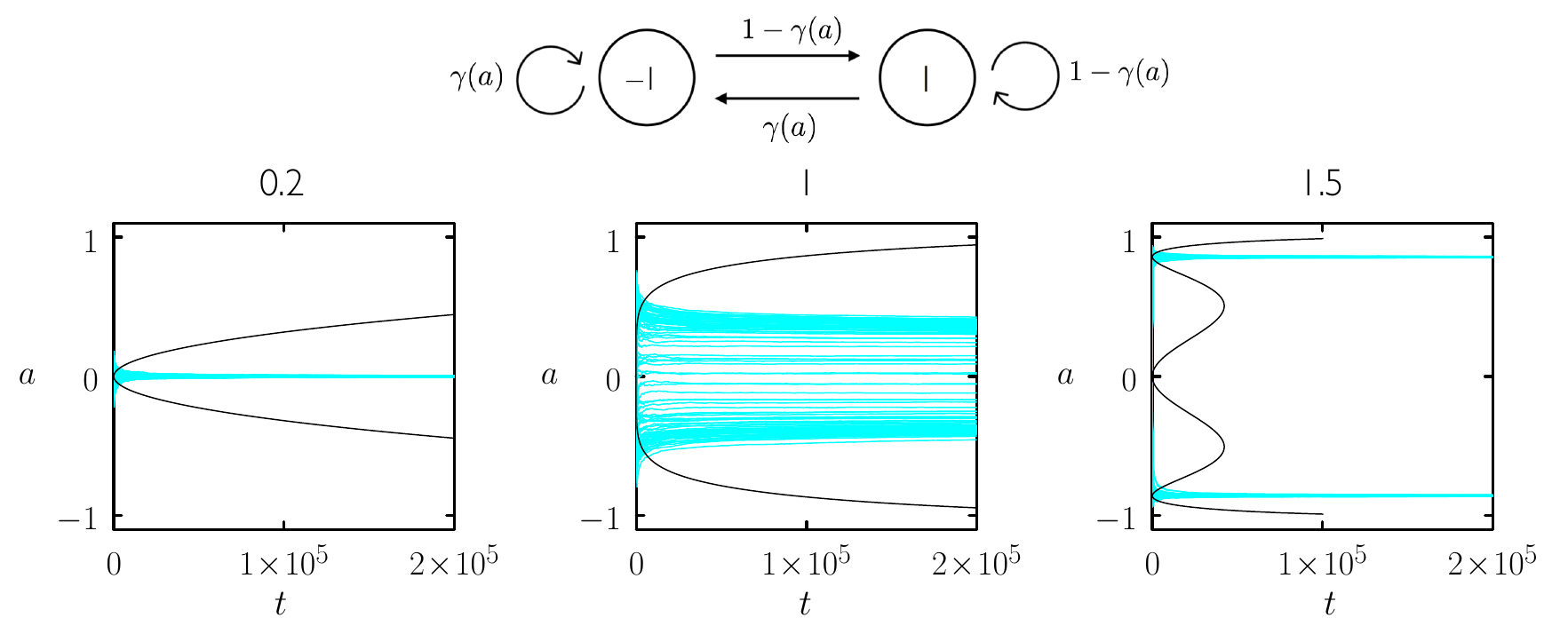} 
   \caption{Time-integrated observable $a$ for ensembles of trajectories of the irreversible growth model of Refs.\c{klymko2017similarity,klymko2017rare}. From left to right we show the one-phase region, the critical point, and the two-phase region. Panels are labeled by the values of the model parameter $J$. The black lines are derived from \eq{eden}.}
\label{fig_grow1}
\end{figure*}

\section{Thermodynamic interpretation of walker models}
\label{thermo}

The random walkers of \s{section_driven} and \s{section_undriven} have a thermodynamic interpretation, if we interpret the time dimension of the walker as a second spatial dimension. Then every trajectory $\omega$ of the walker becomes a configuration of a lattice polymer. On each row $n \in \{1,\dots,N\}$ of the lattice (formerly the time direction), the polymer occupies a single site $x_n \in \{-(L-1)/2,\dots,(L-1)/2\}$. The polymer is held at reciprocal temperature $\beta = (\kt)^{-1}$, and we define the energy of configuration $\omega$ to be
\begin{equation}
  \beta E(\omega) = -\ln \pi_1(x_0^\omega) - \sum_{i=2}^{N}\ln W_{x_{i}^{\omega} x_{i+1}^{\omega}}.
\end{equation}
Here $\pi_1(x)$ is the probability that the polymer has position $x$ on the first row of the lattice.

The probability that the thermodynamic system has configuration $\omega$ is equal to the probability that the dynamical system generates trajectory $\omega$. The probability of the polymer achieving a row-averaged mean position $a$ is
\begin{equation}
  \rho_N(a) = e^{-\beta N g_{N} (a)},
\end{equation}
where 
\begin{equation}
	\beta g_{N}(a) = - \frac{1}{N}\ln \sum_{\omega:a(\omega) = a} e^{-\beta E(\omega)}.
\end{equation}
is the reduced free energy per lattice row. In the limit that $N$ becomes large, the function $g_{N}(a)$ goes over to an $N$ independent function $g(a)$, leading to the large deviation principle 
\begin{equation}\label{define_free_energy}
  \rho_{N}(a) \approx e^{-\beta N g(a)},
\end{equation}
the thermodynamic analog of \eqref{dynamical_LDP}. Taking the Legendre transform of $g(a)$ produces the function $f(k)$ with field $k$. Then $g(a) = I(a)$ and $f(k) = \lambda(k)$. 

As a result, the thermodynamic polymer exhibits the same behavior as the dynamical walker, but does so in space rather than time. In particular, in the case $p=1/2$ the correlation length of the polymer diverges as $L$ diverges, leading to a broadening of the free energy $g(a)$ and the emergence of a kink in $f(k)$. However, analogous to the dynamical case, the probability distribution of polymer positions remains uniform, and no distinct ``phases'' accompany the singularity.

\section{Large-deviation functions of the growth model}
\label{app_grow}

\begin{figure*}[] 
   \centering
  \includegraphics[width=\linewidth]{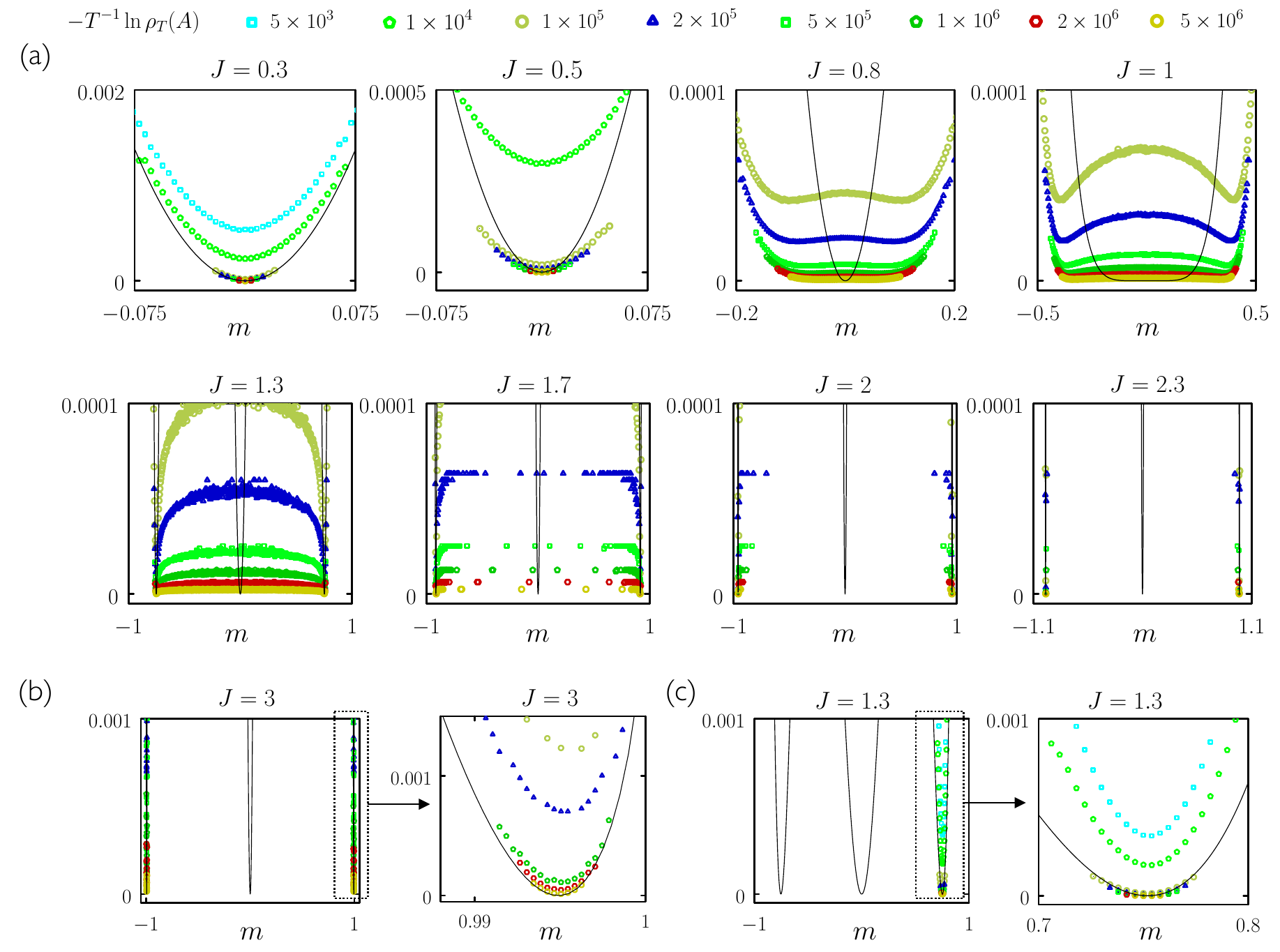} 
   \caption{Empirical rate functions $-T^{-1} \ln \rho_T(A)$ for different values of trajectory time $T$ (denoted by symbols) for the irreversible growth model, compared with the steady-state rate function \eq{eden} (black lines). Panels are labeled by values of the parameter $J$. (a) Far from the critical point $J_{\rm c}=1$ the empirical rate functions are convex in the one-phase region $(J<1)$ and concave in the two-phase region $(J>1)$. Close to the critical point the behavior of the model, on the times simulated, is influenced by a population of transient trajectories. (b) Enlargement of the right-hand portion of the plot for $J=3$ shows the empirical rate function to be consistent with the form \eq{eden}, indicating steady-state growth. (c) We again consider the case $J=1.3$, but now initiate simulations from a pre-made structure of size $T_0=1000$ and composition $a$, consistent with the position of the right-hand minimum of the steady-state rate function \eq{eden}. Doing so allows us to effectively access longer timescales than we could from ``unassembled'' initial conditions. The empirical rate function derived from the resulting ensemble of trajectories is again consistent with the form \eq{eden}.}
\label{fig_erf1}
\end{figure*}

\begin{figure*}[] 
   \centering
  \includegraphics[width=0.9\linewidth]{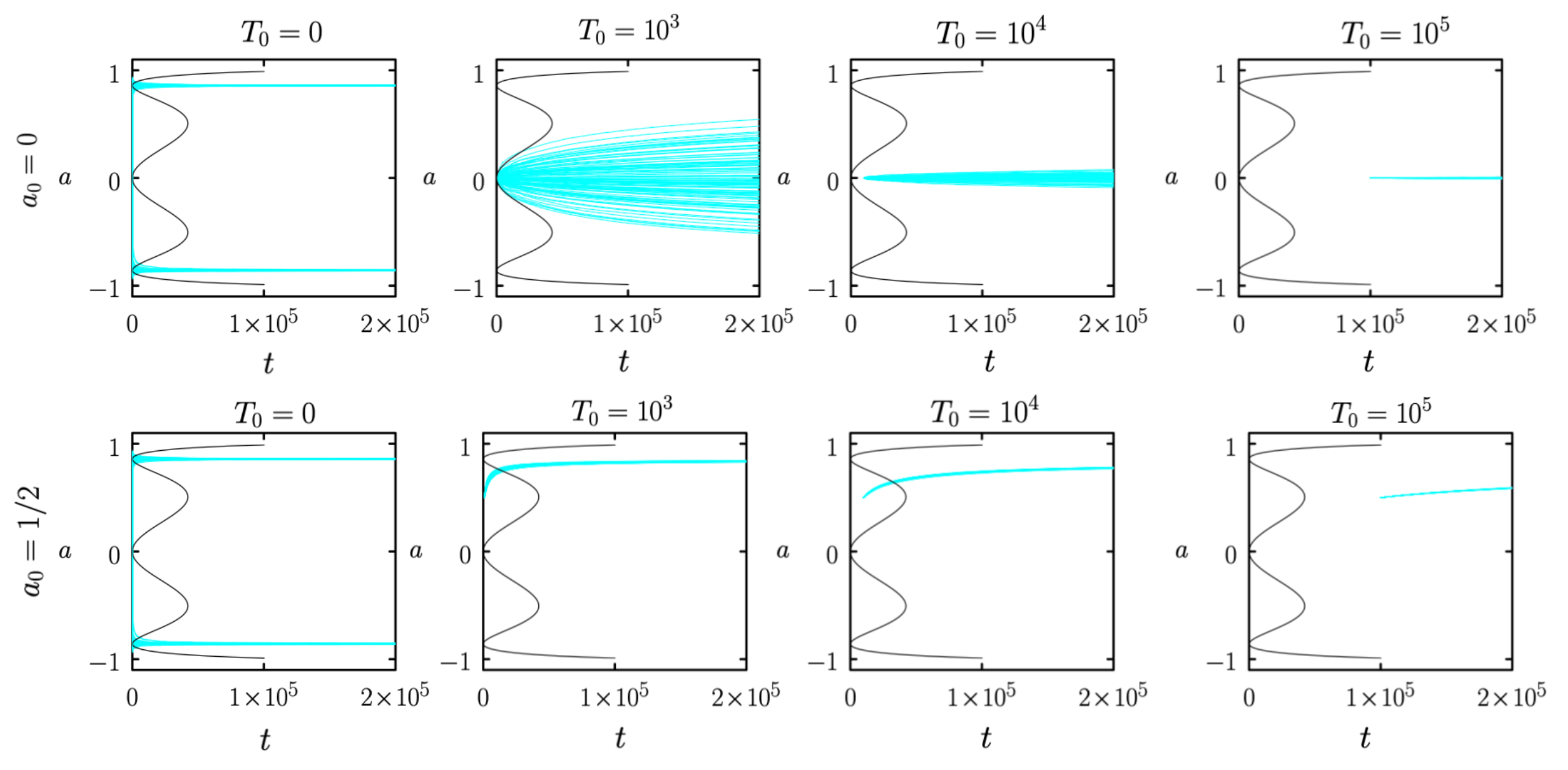} 
   \caption{In the two-phase regime of the growth model, trajectory ensembles begun from structures of increasing size $T_0$, having composition $a_0=0$ (top row), become increasingly strongly confined by the unstable dynamical attractor at $a=0$, corresponding to the central minimum of the function \eq{eden}. By contrast, large structures of initial composition $a_0 \neq 0$ not consistent with any of the minima are driven toward the stable dynamical attractors (in this figure only, the symbol $a_0$ denotes initial composition, and not a typical value of the trajectory ensemble.)}
\label{fig_grow2}
\end{figure*}

The irreversible growth model of Refs.\c{klymko2017similarity,klymko2017rare}, summarized in \s{grow}, is sketched at the top of \f{fig_grow1} in the form of a two-state switch with memory. In that figure we show trajectories of the model at and at either side of the dynamical critical point $J_{\rm c}=1$. Associated with this phase transition is a large-deviation singularity: the rate function for $a$ becomes non-convex at the critical point, and the associated SCGF is kinked\c{klymko2017similarity,klymko2017rare}. The non-convexity in the two-phase region is of different character to that seen in ergodic models that display dynamical intermittency. In this appendix we explore this point further.

The black lines in \f{fig_grow1} are the rate-function bound
\bea
 \label{eden}
I_0(a)&=&  \frac{1-a}{2} \ln \left(1-a\right)+\frac{1+a}{2}\ln \left( 1+a \right)\nonumber \\&-& J a^2 + \ln \cosh Ja,
\eea
derived in \cc{klymko2017rare} under the assumption of steady-state growth. We will call \eqq{eden} the steady-state rate function. It is convex for $J<1$ and non-convex (concave) for $J>1$. 

The steady-state rate function is consistent with the empirical large-deviation rate function of the model in the one-phase and two-phase regions. In \f{fig_erf1} we show empirical rate functions $-T^{-1} \ln \rho_T(A)$ for unbiased trajectories of the growth model, compared with the form \eq{eden}. To measure histograms $\rho_T(A)$ we used of order $10^6$ trajectories, propagated for the times shown, and sampled $a$ using 500 evenly-spaced bins (in \f{fig_erf1}(b) we used 5000 bins and simulation times one-fifth of those in the other panels). Panels are labeled with the value of the coupling $J$, and all simulations (except those in \f{fig_erf1}(c)) were begun from time $T_0=0$, corresponding to an ``unassembled'' structure.

For couplings $J$ well below and well above the critical point, the empirical rate functions are convex and concave, respectively, and are consistent with the steady-state rate function \eq{eden} (see particularly the enlargement of the right-hand side of the plot in \f{fig_erf1}(b)). Near the critical point, on either side, the relaxation time of the model is large\c{klymko2017similarity}, and empirical rate functions, for the times simulated, depart from the form \eq{eden}. (The distribution $\rho_T(A)$ (not the rate function) at the critical point $J_{\rm c}=1$ is bimodal, similar to the magnetization distribution of the 2D critical Ising model in square geometry\c{cardozo2016finite}.) In the two-phase regime, close to the critical point, we can detect by direct simulation a population of transient trajectories that linger for some time near the unstable fixed point $a=0$ (see \f{fig_grow2}), and later commit to one of the stable attractors. These trajectories populate the middle portions of the empirical rate functions close to the critical point.  In this regime the empirical rate function consists of two convex pieces joined by a bar, and in the limit of large $T$ the height of this bar moves to zero.

This behavior is consistent with \cc{jack2019large}, which showed that the rate function $I(a)$ accounting for steady-state {\em and} transient trajectories is zero between the stable minima in the two-phase regime. However, whether the empirical rate function is described by this result or by the steady-state rate function bound \eq{eden} depends strongly upon where in parameter space we operate. Simulations for the times and couplings used in \cc{jack2019large} (e.g. at the point $J=1.3$; see panel in \f{fig_erf1}) are dominated by transient effects, and are not representative of the long-time behavior of the model in the two-phase regime, contrary to the claim made in\c{jack2019large}. As the trajectory time $T$ becomes large, or $J$ is made larger (so reducing the relaxation time of the model), the number of trajectories required to observe transient trajectories. For instance, at $J=3$ and $T > 10^5$, none of $10^8$ trajectories was of the transient type. By contrast, trajectories in the vicinity of the stable attractors can be seen at all times, and the rare trajectories detected by direct simulation result from invasion from those attractors. As a result, for long times and a large but computationally feasible number of trajectories, the empirical rate function of the model in the two-phase region is non-convex (concave), and is consistent with the steady-state rate function \eq{eden}. This concavity reflects ergodicity breaking and the presence of distinct dynamical phases.

The behavior of the growth model compared to that of the walker models reinforces the importance of phenomenology to any classification scheme: these models display similar large-deviation singularities, but support phase transition-like behavior to different extents.


\end{document}